\documentclass[12pt,a4paper,english,superscriptaddress,aps,nofootinbib]{revtex4}
\usepackage[utf8]{inputenc}
\usepackage[T1]{fontenc}
\usepackage{amsmath,amssymb,graphicx}
\makeatletter
\usepackage{babel}
\usepackage[active]{srcltx}
\usepackage{graphicx,color}
\usepackage{changebar}
\usepackage{hyperref}
\usepackage[T1]{fontenc}
\usepackage{esint}
\usepackage{multirow}
\usepackage{dsfont}
\usepackage{ae}
\usepackage{amsmath}
\usepackage{braket}
\usepackage{mathtools}
\usepackage{slashed}
\usepackage{empheq}
\usepackage{multirow}
\usepackage{graphicx}
\usepackage{amsfonts}
\usepackage{amsmath}
\newcommand{\be}{\begin{equation}}
	\newcommand{\ee}{\end{equation}}
\newcommand{\bea}{\begin{eqnarray}}
	\newcommand{\eea}{\end{eqnarray}}

\begin{document}
	\title{Quantum information entropy of a particle trapped by the Aharonov-Bohm-type effect}
	
\author{F. C. E. Lima}
\email{cleiton.estevao@fisica.ufc.br}
\affiliation{Universidade Federal do Cear\'{a}, Departamento do F\'{i}sica, Fortaleza, CE, 60455-760, Brazil.}

\author{A. R. P. Moreira}
\affiliation{Universidade Federal do Cear\'{a}, Departamento do F\'{i}sica, Fortaleza, CE, 60455-760, Brazil.}
 
\author{C. A. S. Almeida}
\email{carlos@fisica.ufc.br}
\affiliation{Universidade Federal do Cear\'{a}, Departamento do F\'{i}sica, Fortaleza, CE, 60455-760, Brazil.}

\author{C. O. Edet}
\email{collinsokonedet@gmail.com}
\affiliation{Institute of Engineering Mathematics, Universiti Malaysia Perlis, 02600 Arau, Perlis, Malaysia.}
\affiliation{Faculty of Electronic Engineering Technology, Universiti Malaysia Perlis, 02600 Arau, Perlis, Malaysia.}
\affiliation{Department of Physics, Cross River University of Technology, Calabar, Nigeria.}

\author{N. Ali}
\email{norshamsuri@unimap.edu.my}
\affiliation{Faculty of Electronic Engineering Technology, Universiti Malaysia Perlis, 02600 Arau, Perlis, Malaysia.}
\affiliation{Advanced Communication Engineering (ACE) Centre of Excellence, Universiti Malaysia Perlis, 01000 Kangar, Perlis, Malaysia.}

\begin{abstract}\vspace{0.7cm}
\noindent \textbf{Abstract:} In this research article, we use the Shannon's formalism to investigate the quantum information entropy of a particle trapped by the Aharonov-Bohm-type field. For quantum information study, it is necessary to investigate the eigenstates of the quantum system, i. e., the wave functions and energies of the quantum states. We assumed that the particle is in principle, confined in a cylindrical box in the presence of Aharonov-Bohm-type effect due to dislocation defect. Analysis of the quantum information entropy, reveals that the dislocation influences the eigenstates and, consequently, the quantum information of the system.

\end{abstract}
\maketitle

\thispagestyle{empty}

\section{Introduction}

Vector and scalar potentials appear in physics as auxiliary tools describing the electric and magnetic fields \cite{AB,Arfken}. Furthermore, potentials play a fundamental role in the canonical description of a physical system at the classical \cite{Marion} and quantum levels \cite{Griffiths,Sakurai}. In quantum theory, potentials are responsible for the emergence of the Aharonov-Bohm (AB) effect \cite{AB}. The AB effect is a quantum phenomenon in which an electrically charged particle is affected by an electromagnetic potential \cite{Griffiths,Sakurai}. In principle, the AB effect tells us that the particle will feel the influence of the field even when the particle is at a null-field region \cite{AB,Griffiths,Sakurai}. This effect was described in 1959 by Yakir Aharonov and David Bohm in their seminal paper {\it Significance of Electromagnetic Potentials in Quantum Theory} \cite{AB}.

Several applications discussing AB effects have appeared in the literature \cite{Overstreet,Ronen,Mefford,FAhmed1,FAhmed2,FAhmed3,Ba1,Ba2,Ba3}. For example, the AB effect arises when studying carbon nanotubes \cite{Ajiki}, in the scattering process \cite{Ruij}, and in photonic systems \cite{Fang}. The fact is that there is a growing interest in the study of the AB effect \cite{Vaidman,Caprez,Chaichian}. In particular, this interest motivates us to understand the quantum information of a particle trapped by an AB-type effect due to the presence of a cosmic string topological defect, i. e., the dislocation defect.

Cosmic strings are topological defects analogous to the flux tubes that arise in type II superconductors and vortex filaments in superfluid helium. The Ref. \cite{Vilenkin} suggests that this defect must have formed in a large unification transition or simply conceivable later in the electroweak transition. It is necessary to mention that the standard model does not predict stable strings. However, some generalizations allow this prediction \cite{Vilenkin}. Generally, the cosmic string is a hypothetical topological defect formed during a symmetry-breaking phase transition in the early universe when the topology of the vacuum manifold associated with this symmetry-breaking was not simply connected. The existence of cosmic string defect was proposed first by Kibble in the seminal paper \textit{topology of cosmic domains and strings} \cite{Kibble,Kibble2}. For more details, see Refs. \cite{Kibble,Kibble2,Helliwell,Gott,Linet}.

The idea of quantum information arises in search of a general communication theory \cite{Shannon}. Claude E. Shannon sought an extension of communication theory to include new factors. In particular, Shannon sought to implement in his study the noise effect on the channel and the possible savings in the propagation process of the message \cite{Shannon}. This relevant study made Shannon one of the forerunners of communication theory.

Shannon's theory has been used in quantum mechanics to study information from physical systems \cite{Collins,LMA,LMMA,Almeida1}. In quantum theory, Shannon's theory is a theoretical measuring of the communication responsible for informing us of the measure of uncertainty in the position and momentum of the particle \cite{LMA,LMMA}. In other words, Shannon's information is information theoretical-measuring associated with the position space and reciprocal. Indeed, this informational analysis plays a relevant role in the foundations of information theory \cite{Shannon} and quantum mechanics problems \cite{Pathria,Griffiths,Sakurai} (see Refs.  \cite{Dong1,Dong2,Dong3}). Furthermore, Shannon's entropy is a useful approach used in the study of cryptography \cite{Grosshans}, noise theory \cite{Wyner}, harmonic oscillators \cite{Boumali,Yanez,Dehesa0}, potential wells \cite{Bouvrie,Torres,Dehesa,Sun1,Majernik,Tserkis,Mukherjee,Kumar,Amadi}, and theories of effective mass  \cite{Serrano,Hua,Navarro,Lima}.

Some studies are found in the literature discussing the influence of the AB effect on the system eigenstates \cite{Ikot0}. Furthermore, some studies were also performed on the information of quantum-mechanical systems with AB effect with a specific interaction \cite{Collins}. However, as far as we know, no studies have been carried out purely studying the influence of the AB-type effect on the quantum information of a spinless particle. Therefore, the main purpose is to study how quantum information changes as the AB-type effect is modified.

We organize the paper as follows: In Sec. II, we build the quantum theory and present the analytical solution of the eigenstates of the system. Posteriorly, in Sec. III, the quantum information of the system is studied considering an alteration of the AB-type effect. Finally, in Sec. IV, we announce our findings.

\section{Quantum description of the system}

Our purpose in this section is the quantum description of a spinless particle subjected to the AB effect in the presence of dislocation. So, the eigenstates depend on the AB effect and the dislocation defect. In summary, the dislocation defect corresponds to the distortion of a vertical line into a spiral \cite{Valanis:2005pc,Puntigam:1996vy,daSilva:2019lzh}. Seeking to understand how the AB effect and dislocation modify the eigenstates, let us consider a quantum system composed of a spinless particle in the presence of a dislocation, i. e.,
\begin{align}\label{1}
    ds^2=dr^2+2\beta\,dr\,d\theta+r^2d\theta^2+dz^2,
\end{align}
where the parameter $\beta$ describes the dislocation and obeys the constraint $0<\beta<1$. Furthermore, the cylindrical symmetry considered (\ref{1}) requires that $0<r<\infty$, $0\leq\theta\leq 2\pi$ and $-\infty\leq z\leq\infty$.

One classifies the topological defects into three classes, i.e., domain walls, cosmic strings, and monopoles, see Ref. \cite{Vachaspati0}. In this article, metric (\ref{1}) describes a cosmic string defect. In particular, cosmic strings are linear spacetime topological defects \cite{Vachaspati0,Helliwell,Gott,Linet,Kibble,Kibble2,Vilenkin}. Physically, we characterize this class of defects by a conical singularity determined by a concentrated curvature on the symmetry axis. Due to the relevant physical repercussions of concepts related to this defect class, several studies have considered cosmic string defects in their investigations. For more details, see Refs. \cite{Ramberg,WAhmed,HChen,KBakkeH,BoumaliNM,HassanabadiHM,ZWang}.

Considering a nonrelativistic particle in the presence of dislocation, we write the Schr\"{o}dinger equation as
\begin{align}\label{Schr}
    -\frac{1}{2m}\frac{1}{\sqrt{g}}\partial_i(g^{ij}\sqrt{g}\partial_j)\psi(r,\theta,z)=\mathcal{E}\psi(r,\theta,z).
\end{align}
Here, we use the $\hbar=c=1$.

Now using the line element (\ref{1}) and considering the Eq. (\ref{Schr}), one obtains
\begin{align}\label{Schr1}
    -\frac{1}{2m}\bigg\{\frac{\partial^2}{\partial r^2}+\frac{\partial}{\partial z^2}+\frac{1}{(r^2-\beta^2)}\bigg[r\frac{\partial}{\partial r}+\bigg(\frac{\partial}{\partial\theta}-\beta\frac{\partial}{\partial z}\bigg)^2\bigg]\bigg\}\psi(r,\theta,z)=\mathcal{E}\psi(r,\theta,z).
\end{align}

To solve Schr\"{o}dinger's equation (\ref{Schr1}), allow us assume 
\begin{align}\label{tr}
    \psi(r,\theta,z)=\text{e}^{i(l\theta+kz)}R(r).
\end{align}
Here, $l=0,\pm 1,\pm 2,\pm 3,...$ and $k\in \mathbb{R}$. Besides, note that the parameters $l$ and $k$ are the eigenvalues associated with the operators $\hat{p}_z=-i\partial_z$ and $\hat{L}_z=-i\partial_\theta$, respectively.

Assuming the transformation (\ref{tr}) where $\theta$ and $z$ are cyclic coordinates, we write Schr\"{o}dinger's equation as
\begin{align}\label{Schr2}
 R''(r)+ \frac{1}{(r^2-\beta^2)}\bigg[rR'(r)-(l-\beta k)^2R(r)\bigg] +(2m\mathcal{E}-k^2)R(r)=0.
\end{align}
Here, prime notation represents the derivative concerning the radial coordinate.

Seeking to describe the confinement of the spinless particle, let us assume the coordinate change
\begin{align}\label{2}
 r=\frac{x}{\sqrt{2m\mathcal{E}-k^2}}.
\end{align}
This coordinate change appears in Refs. \cite{daSilva:2019lzh,MulerK} to study a particle confinement by a hard wall.

Considering the Eq. (\ref{2}), we rewrite the Eq. (\ref{Schr2}) as
\begin{align}\label{Schr3}
 \bigg(1-\frac{4\varsigma^2}{x^2}\bigg)\ddot{R}(x)+\frac{1}{x}\dot{R}(x)+\bigg(1-\frac{\lambda}{x^2}-\frac{2\varsigma^2}{x^2}\bigg)R(x)=0,
\end{align}
where 
\begin{align}
    \lambda=(l-\beta k)^2+\varsigma^2,
\end{align}
and
\begin{align}
    \varsigma^2=\frac{\beta^2(2m\mathcal{E}-k^2)}{4}.
\end{align}
Here, the point notation represents the derivative concerning the variable $x$.

Let us highlight that the parameter that describes the dislocation (or distortion of the metric) has a value restrained in the range $[0,1]$ \cite{Valanis:2005pc,Puntigam:1996vy}. So, one can assume the parameter $\beta$ is small. Therefore, the higher order terms of $\beta$ will be negligible. In this way, we reformulate the Eq. (\ref{Schr3}) as follows:
\begin{align}\label{Schr4}
 \ddot{R}(x)+\frac{1}{x}\dot{R}(x)+\bigg[1-\bigg(\frac{l-\beta k}{x}\bigg)^2\bigg]R(x)=0.
\end{align}

Here, allow us to remember that every equation written in the form
\begin{align}\label{bessel}
    f''(x)+\frac{1}{x}f'(x)+\bigg(k^2-\frac{\alpha^2}{x^2}\bigg)f(x)=0
\end{align}
is known as Bessel's equation \cite{Arfken,Abramowitz,Butkov}. Thus, the general solution of the Bessel equation will be
\begin{align}\label{generalSBessel}
    f(x)=c_1\,J_\alpha(x)+c_2\,J_{-\alpha}(x),
\end{align}
where
\begin{align}\label{SBessel}
    J_{\alpha}(x)=\sum_{k=0}^{\infty}(-1)^k\frac{1}{k!\,\Gamma(\alpha+k+1)}\bigg(\frac{x}{2}\bigg)^{\alpha+2k}.
\end{align}
For more details on the Bessel equation, see Refs. \cite{Arfken,Abramowitz,Butkov}.

Comparing Eq. (\ref{bessel}) and its solution (\ref{generalSBessel}) with Eq. (\ref{Schr4}) and requiring the wave function to be normalizable, one obtains that the solution of Eq. (\ref{Schr4}) is
\begin{align}\label{3}
 R(x)=A_0J_{|l-\beta k|}(x).
\end{align}
Here $A_0$ is the normalization constant, and $J_{|l-\beta k|}(x)$ is Bessel's function of first kind \cite{Arfken,Abramowitz,Butkov}.

Building the general wave function that describes spinless particles in the presence of disclination, one obtains
\begin{align}\label{4}
\psi(r,\theta,z)=A_0\text{e}^{i(l\theta+k z)}J_{|l-\beta k|}\bigg(\frac{2\varsigma}{\beta^2}r\bigg).
\end{align}

As discussed by Silva and Bakke \cite{daSilva:2019lzh}, let us analyze the confinement of the spinless particle by a hard-wall confining potential in the elastic medium. For this, allow us to assume that
\begin{align}\label{cond0}
    \lim_{x\to x_0}R(x)=0 \, \, \, \, \, \, \text{with} \, \, \, \, \, \, x_0=\beta\sqrt{2m\mathcal{E}-k^2}.
\end{align}

Applying the constraint (\ref{cond0}), one obtains that 
\begin{align}\label{6}
\mathcal{E}_{n,l,k}=\frac{1}{2m}\bigg(\frac{\Theta_{n,\vert l-\beta k\vert}}{r_0}\bigg)^2+\frac{k^2}{2m}.
\end{align}
Here $\Theta_{n,\vert l-\beta k\vert}$ is the $n$-th zero of the $\vert l-\beta k\vert$-th Bessel's function \cite{Griffiths}. Furthermore, $n=0, 1, 2, \dots$, and the quantum number associated to the angular momentum is $l=0,\pm 1,\pm 2, \dots$. 
 
The results of the Eqs. (\ref{4}) and (\ref{6}) expose how eigenstates change with disclination. So, we expect the existence of a dependency on the quantum information with the disclination defect. In the next section, we present a study of the variation of Shannon information as the disclination changes.

Note that the parameters $n$ and $\vert l-\beta k\vert$ play the role of quantum numbers associated with the radial and angular variables, respectively. Therefore, this result allows us to perceive the influence of the defect on the energy eigenvalues. Moreover, one notes this influence by a change in the quantum number of the angular momentum. It is relevant to mention that this change arises even if there is no interaction between the particle and the defect. Indeed, this effect is called the Aharonov-Bohm-type effect and appears due to the presence of the topological defect.% \cite{daSilva:2019lzh,SNCF}.

\section{Information theoretical measure: Shannon's entropy}

The growing interest in studies of information-theoretical measurements of quantum-mechanical systems is evident \cite{Grosshans,Wyner,Boumali,Yanez,Dehesa0,Bouvrie,Torres,Dehesa,Sun1,Majernik,Tserkis,Mukherjee,Kumar,Amadi,Serrano,Hua,Navarro,Lima,EI}. Initially, the entropic measure of information comes up with an alternative proposal to the well-known Heisenberg uncertainty relation \cite{Navarro,BBM}. Among different approaches used to measure information entropies is the so-called Shannon's entropy \cite{Navarro}. In quantum theories, Shannon's entropy plays an essential role in uncertainty measurement. In particular, this entropy gives us a new version of Heisenberg's uncertainty relationship between the position and momentum spaces \cite{BBM}. This relation is
\begin{align}
    S_\textbf{r}+S_\textbf{p}\geq D(1+\text{ln}\pi).
\end{align}
Here, $D$ is the spatial dimension (in our case $D=3$), $S_\textbf{r}$ is the information entropy at the position space, and $S_\textbf{p}$ is the information entropy at the momentum space. 

Using the definition of Shannon's entropy, let us write the quantum information entropy as
\begin{align}
    \label{Sr}
    S_{\textbf{r}}=-\int_{-\infty}^{\infty}\int_{0}^{2\pi}\int_{0}^{\infty}\,\vert\psi(r,\theta,z)\vert^2\,\text{ln}\, \vert\psi(r,\theta,z)\vert^2\, r\,dr d\theta dz,
\end{align}
and
\begin{align}
    \label{Sp}
    S_{\textbf{p}}=-\int_{-\infty}^{\infty}\int_{0}^{2\pi}\int_{0}^{\infty}\,\vert\psi(p_r,p_\theta,p_z)\vert^2\,\text{ln}\, \vert\psi(p_r,p_\theta,p_z)\vert^2\,p_r\,dp_r dp_\theta dp_z,
\end{align}
where
\begin{align}\label{Fourier}
    \psi(p_r,p_\theta,p_z)=&\int_{0}^{\infty}r\, dr\int_{0}^{2\pi}d\theta\int_{-\infty}^{\infty}\,dz\,\psi(r,\theta,z)\,\text{e}^{-i2\pi p_r r\cos(p_\theta-\theta)}\,\text{e}^{-i2\pi p_z z}.
    %\frac{1}{\sqrt{2\pi}}\int_{-\infty}^{\infty}\, \psi(x_i)\,\text{e}^{ip_ix_i}\, dx_i,
\end{align}
i. e., the three-dimensional Fourier transform in cylindrical coordinates. For more details, see Refs. \cite{Arfken,Abramowitz,Butkov}.

Shannon's entropy helps us to calculate the entropic information measures. In this context, Shannon's entropy has performed a relevant role in several studies. For example, one can apply this approach in investigating Aharonov-Bohm rings \cite{Collins}, quantum systems with double-well potentials \cite{Dong2}, hyperbolic interaction \cite{Torres}, position-dependent mass theories \cite{Lima,Hua}, and the Aharonov-Bohm effect \cite{CI}. 

\begin{figure}[ht!]
\begin{center}
\includegraphics[height=6cm,width=7.5cm]{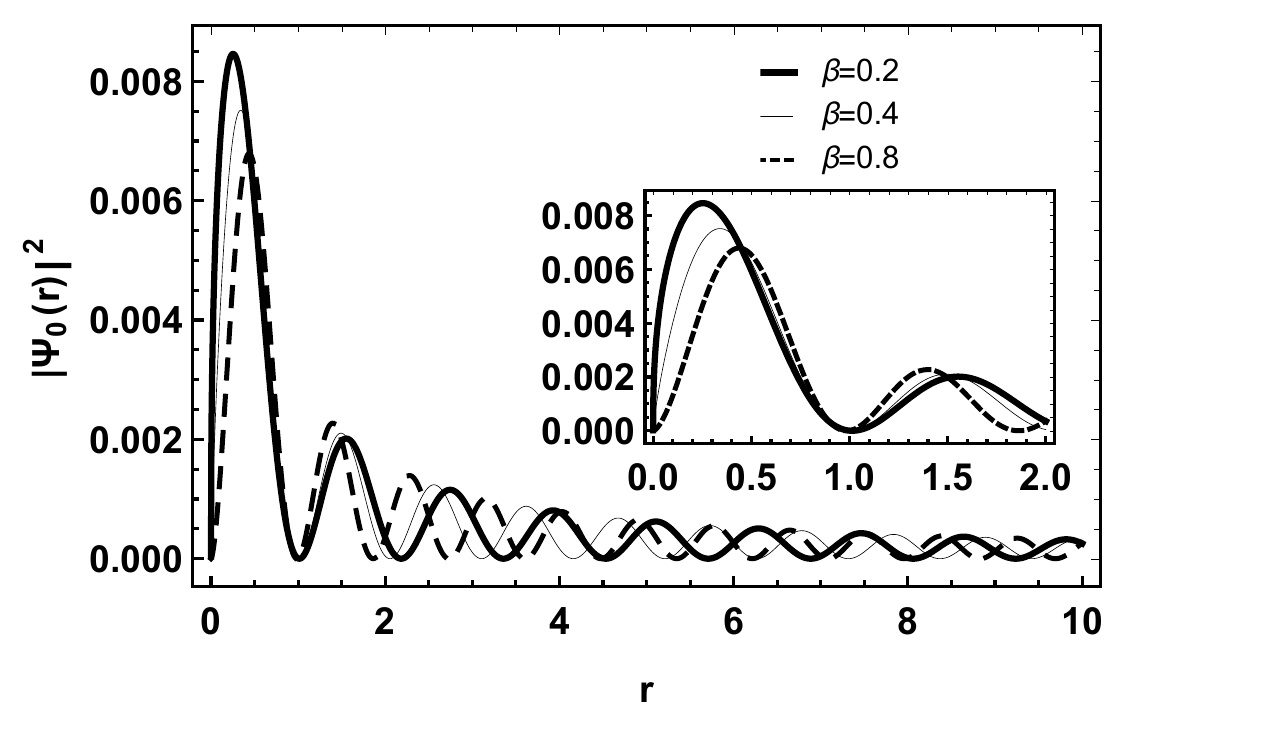}
\includegraphics[height=6cm,width=7.5cm]{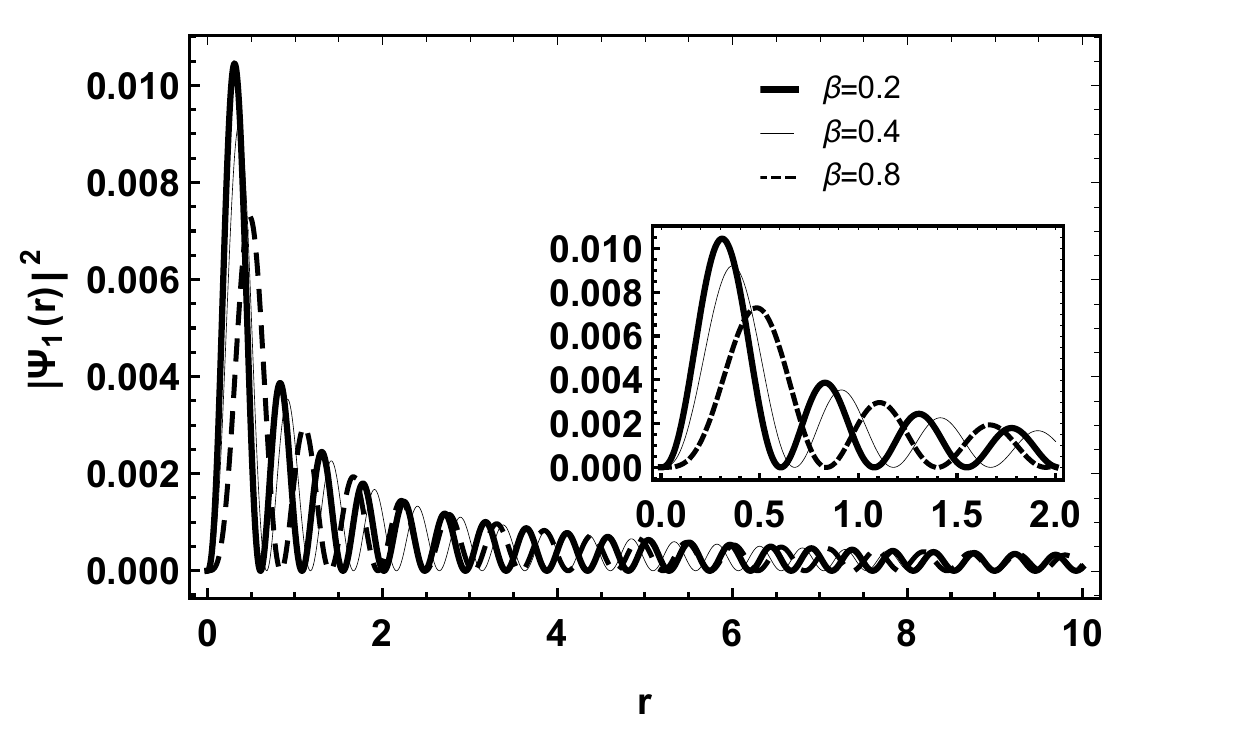}\\ \vspace{-0.25cm}
    (a) \hspace{7cm} (b)\\
\includegraphics[height=6cm,width=7.5cm]{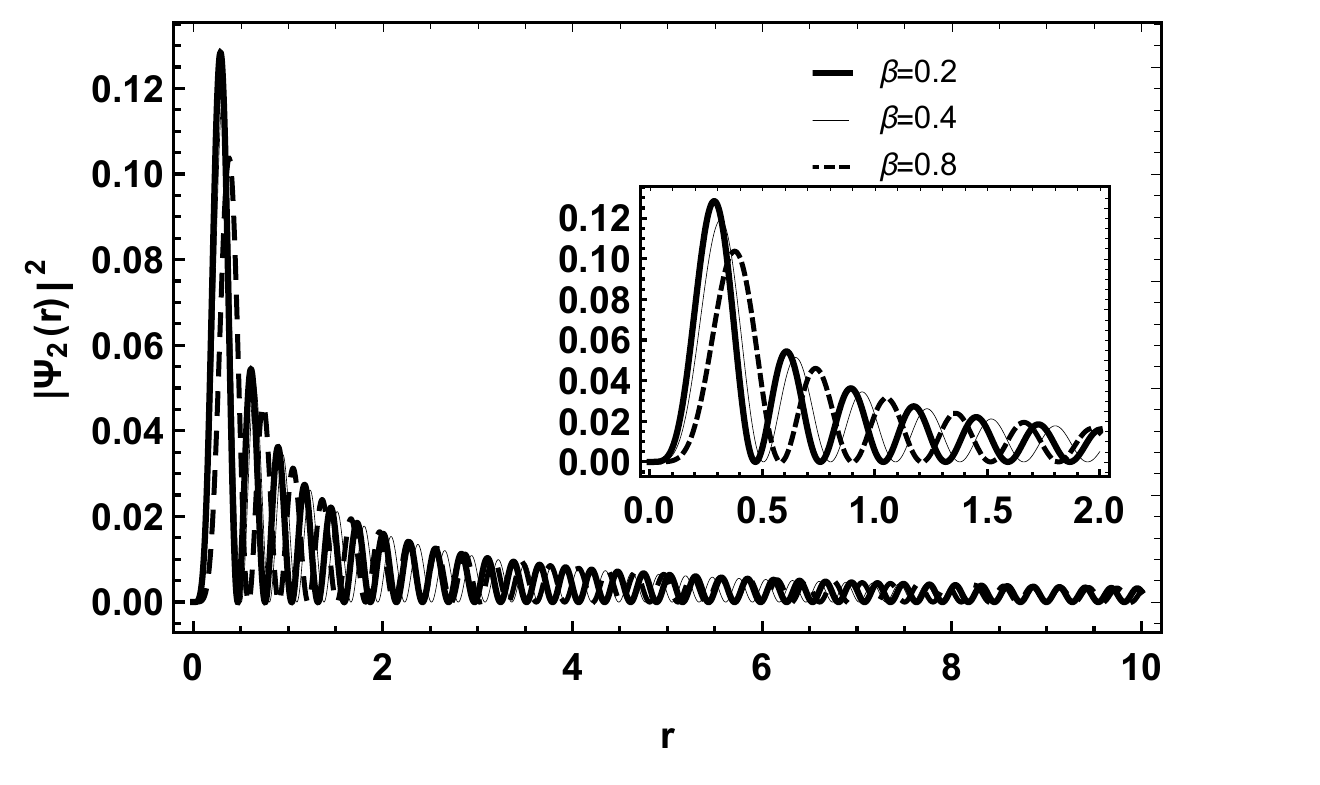}\\ \vspace{-0.25cm}
(c)
\end{center}\vspace{-0.5cm}
\caption{Probability density at the position space. (a) The case: $n=0$ and $l=0$. (b) The case: $n=1$ and $l=-1$. (c) The case: $n=2$ and $l=-2$.
\label{fig1}}
\end{figure}

\begin{figure}[ht!]
\begin{center}
\includegraphics[height=6cm,width=7.5cm]{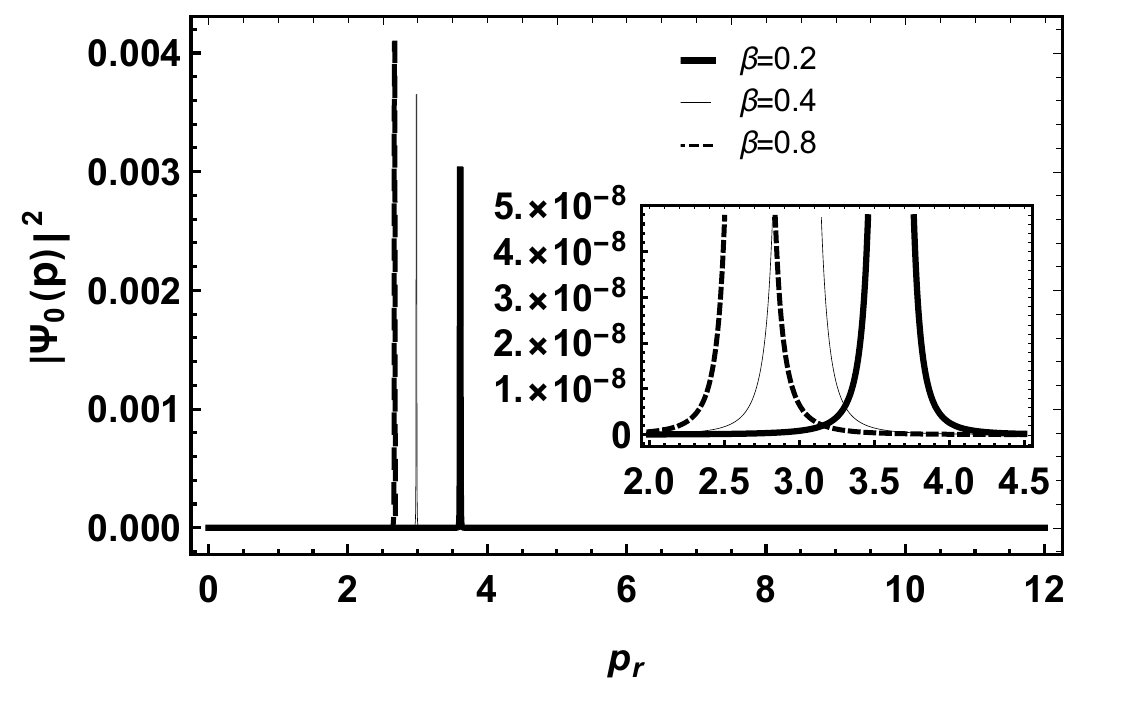}
\includegraphics[height=6cm,width=7.5cm]{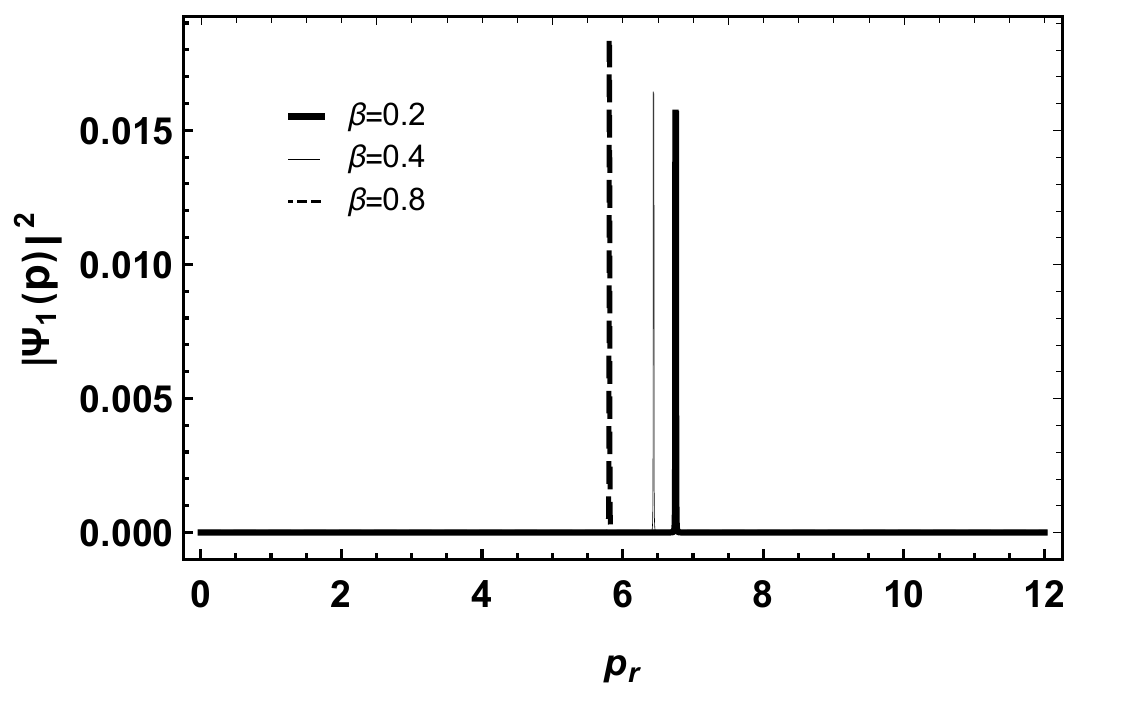}\\ \vspace{-0.25cm}
 \hspace{0.3cm} (a) \hspace{7cm} (b)\\
\includegraphics[height=6cm,width=7.5cm]{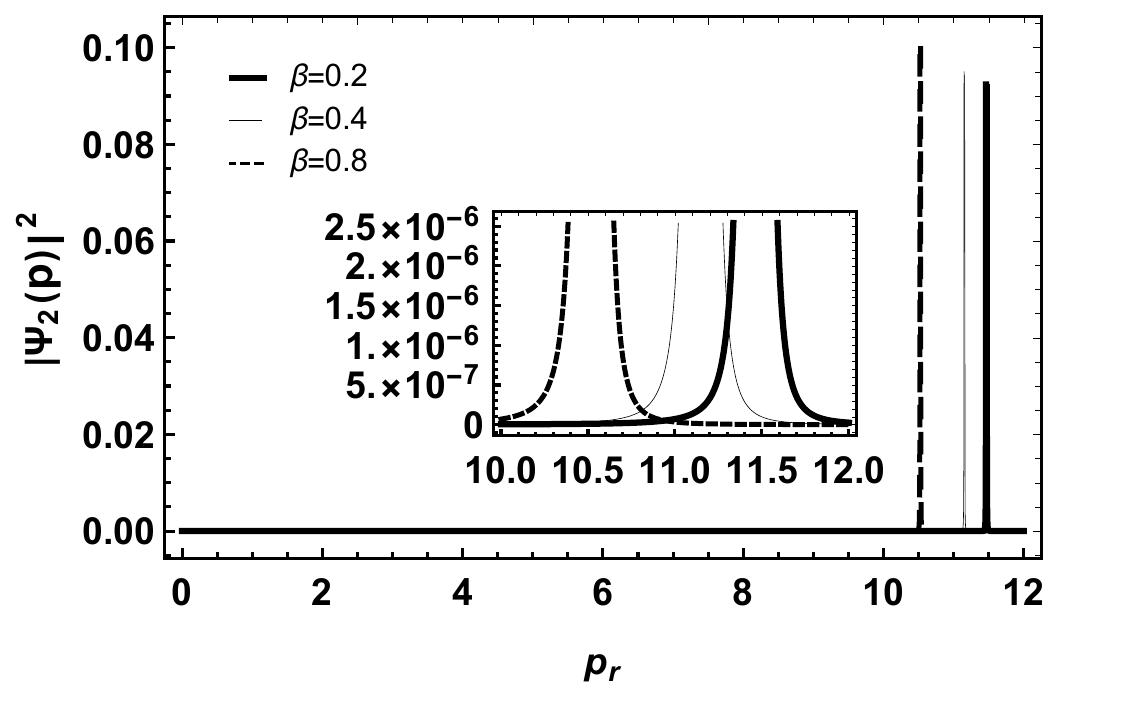}\\ \vspace{-0.25cm}
 (c)
\end{center}\vspace{-0.5cm}
\caption{Probability density at the reciprocal space. (a) The case: $n=0$ and $l=0$. (b) The case: $n=1$ and $l=-1$. (c) The case: $n=2$ and $l=-2$.
\label{fig2}}
\end{figure}

Motivated by these applications, we will use this approach to study quantum information theoretical measurements of a spinless particle. That particle is in the presence of a screw displacement, which produces the Aharonov-Bohm-type effect. It is important to mention that the study of theoretical measurements of information of this system is the first step to understanding the gain (or loss) of quantum information of particles that interact with this class of topological defect. So, to achieve our purpose, let us consider the wave eigenfunctions (\ref{4}) and substitute in Eq. (\ref{Sr}). This substitution leads us to Shannon's entropy in the position space. Subsequently, we perform Fourier's transform (\ref{Fourier}) of the eigenfunctions (\ref{4}) and calculate Shannon's entropy in the reciprocal space using Eq. (\ref{Sp}). The numerical results shown in the table \ref{Ttab1} are obtained numerically solving the integrals (\ref{Sr}) and (\ref{Sp}) for the first eigenstates. In Fig. \ref{fig1}, the probability densities at the position space for the ground state and the first two excited states used in Shannon's entropy calculation are displayed. On the other hand, in Fig. \ref{fig2}, probability densities at the reciprocal space used in Shannon's entropy calculation are exposed.

Analyzing Fig. \ref{fig1}, i. e., the probability density for the first three energy levels, one notes that the more energetic states have greater amplitudes and oscillations. In addition, the results presented in Fig. \ref{fig1} show that by increasing the influence of the topological defect responsible for the Aharonov-Bohm type effect, the amplitudes of the probability density increase. That suggests that the quantum information of the system increases as the Aharonov-Bohm-type effect increases (i.e., as the parameter $\beta$ increases).

Looking at Fig. \ref{fig2}, one can see that the probability density of the particle at the reciprocal space is a Dirac delta-like distribution. This probability density profile suggests minimal uncertainty in the momentum of the spinless particle. Furthermore, with the probability densities shown in Figs. \ref{fig1} and \ref{fig2}, we obtain the numerical results of the table \ref{Ttab1}. This result tells us that the quantum information decreases as the disclination increases. This decreasing in information suggests an increase in uncertainty measurements associated with particle position measurements. Besides, the quantum information at the reciprocal space increases when the disclination is larger. Thus, suggesting less uncertainty in the momentum measurement of the particle. Furthermore, one notes that these results remain valid for quantum states with higher energy. It is worth mentioning that in all cases, we have
\begin{align}
    S_r+S_p\geq 6.43419,
\end{align}
and therefore, the Bialynicki-Birula and Mycielski \cite{BBM} relationship is preserved.

\begin{table}[!ht]
\centering
\resizebox{6cm}{6.5cm}{%
\begin{tabular}{|c|c|c|c|c|c|c|c|}\hline
$n$ & $l$ & $\beta$ & $S_{\textbf{r}}$ & $S_{\textbf{p}}$ & $S_{\textbf{r}}+S_{\textbf{p}}$ & $3(1+\ln\pi)$\\ \hline
  &   & 0.2   & 9.74631  & 0.06678   & 9.81309& \\
0 & 0 & 0.4  & 9.74262 & 0.07558  & 9.81821& 6.43419 \\
  &   & 0.8 & 9.74040 & 0.12158  & 9.86199 &\\ \hline
  &   & 0.2   & 9.74435  & 0.05526  & 9.79961& \\
  &-1 & 0.4  & 9.74424 & 0.10408  & 9.84832& 6.43419 \\
  &   & 0.8 & 9.74387 & 0.29596 & 10.03984 &\\ \cline{2-7}
  &   & 0.2   & 9.74483  & 0.00770  & 9.75254& \\
1 & 0 & 0.4  & 9.74439 & 0.10381 & 9.84821& 6.43419\\
  &   & 0.8 & 9.74353 & 0.29641  & 10.03991& \\ \cline{2-7}
  &   & 0.2   & 9.74410  & 0.01640  & 9.76050& \\
  & 1 & 0.4  & 9.74377 & 0.13777  & 9.88154 & 6.43419\\
  &   & 0.8 & 9.74312 & 0.29938  & 10.04251 & \\ \hline
  &   & 0.2   & 9.74461  & 0.02251  & 9.76713& \\
  &-2 & 0.4  & 9.74329 & 0.22803  & 9.97133& 6.43419 \\
  &   & 0.8 & 9.74277 & 0.35949  & 10.10231 &\\ \cline{2-7}
  &   & 0.2   & 9.74518  & 0.02720  & 9.77238& \\
  & -1 & 0.4  & 9.74361 & 0.25583  & 9.99944& 6.43419\\
  &   & 0.8 & 9.74301 & 0.48609  & 10.22910& \\ \cline{2-7}
  &   & 0.2   &9.74482  & 0.04111  & 9.78593& \\
2  & 0 & 0.4  & 9.74370 & 0.25717  & 10.00091 & 6.43419\\
  &   & 0.8 & 9.74317 & 0.52424  & 10.26742 & \\ \cline{2-7}
  &   & 0.2   & 9.74435  & 0.04188  & 9.78623& \\
  & 1 & 0.4  & 9.74335 & 0.35605  & 10.09942& 6.43419 \\
  &   & 0.8 & 9.74291 &0.85920  & 10.60214 &\\ \cline{2-7}
  &   & 0.2   & 9.74399  & 0.07462  & 9.81861& \\
  & 2 & 0.4  & 9.74311 & 0.44406  & 10.18722& 6.43419\\
  &   & 0.8 & 9.74272 & 0.91082  & 10.65351& \\ \hline
\end{tabular}}
\caption{Numerical result of Shannon's entropy.}
\label{Ttab1}
\end{table}

Finally, analyzing the numerical results presented in table \ref{Ttab1}, one notes that the quantum information entropy at the position space is always greater than at the momentum space. That is because the probability regions at the position space (Fig. \ref{fig1}) are larger than the probability regions at the momentum space (Fig. \ref{fig2}). Thus, the uncertainty measure related to the particle position is always greater. Furthermore, we notice that the quantity $S_\textbf{r}+S_\textbf{p}$ decreases when the topological defect contribution decreases. That is because, in the absence of the topological defect, we recover the usual problem of a free particle in a space with cylindrical symmetry. So, in this limit, the particle's uncertainty measures in both spaces are minimal once the cosmic string defects (dislocation) do not contribute in this case. Meantime, the uncertainty measures on the position and momentum increase as the states become more energetic, i.e., when $n$ and $l$ increase. That is a consequence of the particles, in this case, being in more energetic states when $n$ and $l$ increase. In other words, the particles have a momentum increase, and their measurement uncertainties concerning their velocity and position suffer alterations. These results are very interesting and allow us to conclude that Heisenberg's uncertainty principle remains preserved in our theory. To finish, we observe that the particle energy one modifies for different values of $l$ ($l=2$ and $l=-2$). Thus, the uncertainty measures related to the position and momentum of the particle will be distinct.

\section{Final remarks}
In this research, we have employed   Shannon's entropy to study the quantum information entropy of a non-relativistic particle trapped by the Aharonov-Bohm-type effect. The Schrodinger equation in cylindrical coordinate is solved in the presence of topological defect and the resulting Aharonov-Bohm effect to obtain the wave functions and eigenvalues of quantum system. These eigensolutions are employed to evaluate the Shannon entropy numerically. From our analysis of the quantum information, it is possible to seen that the dislocation influences the eigenstates and, consequently, the quantum information of the system. It is interesting to point out that the famous Bialynicki-Birula and Mycielski (BBM) relationship is satisfied as well. Going forward, the results of this study will find applications in quantum information processing and related areas.

\section*{Authors Declaration}

\subsection*{Conflicts of interest/Competing interest}

All the authors declared that there is no conflict of interest in this manuscript.

\subsection*{Acknowledgements}

F. C. E. Lima is grateful to Coordena\c{c}\~{a}o de Aperfei\c{c}oamento de Pessoal de N\'{i}vel Superior (CAPES), n$\textsuperscript{\underline{\scriptsize o}}$ 88887.372425/2019-00. C. A. S. Almeida thanks to Conselho Nacional de Desenvolvimento Cient\'{\i}fico e Tecnol\'{o}gico (CNPq), n$\textsuperscript{\underline{\scriptsize o}}$ 309553/2021-0. C. O. Edet acknowledges eJDS (ICTP). This research was performed partially under LRGS Grant LRGS/1/2020/UM/01/5/2 (9012-00009) Fault Tolerant Photonic Quantum States for Quantum Key Distribution Provided by the Ministry of Higher Education Malaysia (MOHE).

\section*{Data Availability}
One can find the datasets that support the findings of this study through the article.


\begin{thebibliography}{99}

\bibitem{AB}
Y. Aharonov and D. Bohm, Phys. Rev. {\bf 115} (1959) 485.

\bibitem{Arfken}
G. B. Arfken and H. J. Weber, {\it Mathematical methods for physicists}, (6th ed., Elsevier Academic Press, New York, 2005).

\bibitem{Marion}
J. B. Marion, {\it Classical dynamics of particles and systems}, (5th Revised ed., Cengage Learning, USA, 2003).

\bibitem{Griffiths}
D. J. Griffiths, \textit{Introduction to quantum mechanics}, (Prentice Hall, USA, New Jersey, 1994).

\bibitem{Sakurai}
J. J. Sakurai and J. Napolitano, {\it Modern quantum mechanics}, (Addison-Wesley Publishing Company, USA, 1994).

\bibitem{Overstreet}
C. Overstreet, P. Asenbaum, J. Curti, M. Kim and M. A. Kasevich, Science {\bf 375} (2022) 226.

\bibitem{Ronen}
Y. Ronen, T. Werkmeister, D. H. Najafabadi, A. T. Pierce, L. E. Anderson, Y. J. Shin, S. Y. Lee, Y. H. Lee, B. Johnson, K. Watanabe, T. Taniguchi, A. Yacoby and P. Kim, Nature \textbf{16} (2021) 563.

\bibitem{Mefford}
E. Mefford and K. Suzuki, JHEP \textbf{05} (2021) 026.

\bibitem{FAhmed1}
F. Ahmed, Europhys. Lett. \textbf{130} (2020) 40003.

\bibitem{FAhmed2}
F. Ahmed, Advances in High Energy Physics \textbf{2020} (2020) 1-10.

\bibitem{FAhmed3}
F. Ahmed, Europhys. Lett. \textbf{131} (2020) 30002.

\bibitem{Ba1}
A. V. D. M. Maia and K. Bakke, Quant. Stud.: Mathematics and Foundations \textbf{10} (2023) 79.

\bibitem{Ba2}
K. Bakke, Int. J. Mod. Phys. A \textbf{36} (2021) 2150136.

\bibitem{Ba3}
S. L. R. Vieira and K. Bakke, Phys. Rev. A \textbf{101} (2020) 032102.

\bibitem{Ajiki}
H. Ajiki and T. Ando, Physica B \textbf{201} (1994) 349.

\bibitem{Ruij}
S. N. M. Ruijsenaars, Ann. Phys. \textbf{146} (1983) 1-34.

\bibitem{Fang}
K. Fang, Z. Yu and S. Fan, Phys. Rev. Lett. \textbf{108} (2012) 153901.

\bibitem{Vaidman}
L. Vaidman, Phys. Rev. A {\bf 86} (2012) 040101(R).

\bibitem{Caprez}
A. Caprez, B. Barwick and H. Batelaan, Phys. Rev. Lett. \textbf{99} (2007) 210401.

\bibitem{Chaichian}
M. Chaichian, P. Pre\v{s}najdera, M. M. Sheikh-Jabbaric and A. Tureanu, Phys. Lett. B \textbf{527} (2002) 149.

\bibitem{Vilenkin}
A. Vilenkin, Phys. Rev. D \textbf{23} (1981) 852.

\bibitem{Kibble}
T. W. B. Kibble, J. Phys. A \textbf{9} (1976) 1387.

\bibitem{Kibble2}
T. W. B. Kibble, Phys. Rep. \textbf{67} (1980) 183.

\bibitem{Helliwell}
T. M. Helliwell and D. A. Konkowski, Am. J. Phys. \textbf{55} (1987) 401.

\bibitem{Gott}
J. R. Gott, Ap. J. \textbf{288} (1985) 422.

\bibitem{Linet}
B. Linet, Gen. Rel. Grav. \textbf{17} (1985) 1109.

\bibitem{Shannon}
C. E. Shannon, Bell Syst. Tech. J. {\bf 27} (1948) 623.

\bibitem{Collins}
C. O. Edet, F. C. E. Lima, C. A. S. Almeida, N. Ali and M. Asjad, Entropy, {\bf 24} (2022) 1059.

\bibitem{LMA}
F. C. E. Lima, A. R. P. Moreira and C. A. S. Almeida, Int. J. Quant. Chem. {\bf 121} (2021) e26645.

\bibitem{LMMA}
F. C. E. Lima, A. R. P. Moreira, L. E. S. Machado and C. A. S. Ameida, Int. J. Quant. Chem. {\bf 121} (2021) e26749.

\bibitem{Almeida1}
C. A. S. Almeida, C. O. Edet, F. C. E. Lima, N. Ali and M. Asjad, Results in Physics \textbf{47} (2023) 106343.

\bibitem{Pathria}
R. K. Pathria, \textit{Statistical Mechanics}, (2nd ed., Butterworth Heinemann, Jordan Hill, Oxford,  1996).

\bibitem{Dong1}
C. A. Gil-Barrera, R. S. Carrillo, G.-H. Sun and S. -H. Dong, Entropy \textbf{24} (2022) 604.

\bibitem{Dong2}
G. -H. Sun, S. -H. Dong, K. D. Launey,T. Dytrych and J. P. Draayer, Int. J. Quant. Chem. \textbf{115} (2015) 891.

\bibitem{Dong3}
G. -H. Sun,S. -H. Dong and N. Saad, Int. J. Quant. Chem. \textbf{525} (2013) 934.

\bibitem{Grosshans}
F. Grosshans, N. J. Cerf, Phys. Rev. Lett. {\bf 92} (2004) 047905.

\bibitem{Wyner}
A. D. Wyner, S. Shamai, Proc. IEEE {\bf 86} (1998) 442.

\bibitem{Boumali}
A. Boumali, M. Labidi, Mod. Phys. Lett. A {\bf 33} (2018) 1850033.

\bibitem{Yanez}
R. J. Ya\~{n}ez, W. V. Assche and J. S. Dehesa, Phys. Rev. A {\bf 50} (1994) 3065.

\bibitem{Dehesa0}
J. S. Dehesa, E. D. Belega, I. V. Toranzo and A. I. Aptekarev, Int. J. Quantum Chem. {\bf 119} (2019) e25977.


\bibitem{Bouvrie}
P. A. Bouvrie, J. C. Angulo, J. S. Dehesa, Phys. A {\bf 390} (2011) 2215.

\bibitem{Torres}
R. Valencia-Torres, G. H. Sun and S. H. Dong, Phys. Scr. {\bf 90} (2015) 035205.

\bibitem{Dehesa}
J. S. Dehesa, A. Martínez-Finkelshtein, V. N. Sorokin, Mol. Phys. {\bf 104} (2006) 613.

\bibitem{Sun1}
G. -H. Sun, M. A. Aoki, S. -H. Dong, Chin. Phys. B {\bf 22} (2013) 050302.

\bibitem{Majernik}
V. Majernik, R. Charvot, E. Majernikova, J. Phys. A: Math. Gen. {\bf 32} (1999) 2207.

\bibitem{Tserkis}
S. T. Tserkis, C. C. Moustakidis, S. E. Massen, C. P. Panos, Phys. Lett. A {\bf 378} (2014) 497.

\bibitem{Mukherjee}
N. Mukherjee, A. K. Roy, Int. J. Quantum Chem. {\bf 11}, 2018, e25596.

\bibitem{Kumar}
P. R. Kumar, G. Kumar, R. Kumar, A. Kumar, Int. J. Quantum Chem. {\bf 116} (2016) 1413.

\bibitem{Amadi}
P. O. Amadi, A. N. Ikot, A. T. Ngiangia, U. S. Okorie, G. J. Rampho, H. Y. Abdullah, Int. J. Quantum Chem. {\bf 120} (2020) e26246.

\bibitem{Serrano}
F. A. Serrano, B. J. Falaye, S. H. Dong, Phys. A Stat. Mech. Appl. {\bf 446} (2016) 152.

\bibitem{Hua}
S. Guo-Hua, D. Popov, O. Camacho-Nieto, S. H. Dong, Chin. Phys. B {\bf 24} (2015) 100303.

\bibitem{Navarro}
G. Yañez-Navarro, G. H. Sun, T. Dytrych, K. D. Launey, S. H. Dong, J. P. Draayer, Ann. Phys. {\bf 348} (2014) 153.

\bibitem{Lima}
F. C. E. Lima, Ann. Phys. \textbf{442} (2022) 168906.

\bibitem{EI}
C. O. Edet and A. N. Ikot, Eur. Phys. J. Plus \textbf{136} (2021) 432.

\bibitem{Ikot0}
A. N. Ikot, C. O. Edet, P. O. Amadi, U. S. Okorie, G. J. Rampho and H. Y. Abdullah, Eur. Phys. J. D \textbf{74} (2020) 159.

\bibitem{Valanis:2005pc}
K. S. Valanis and V. P. Panoskaltsis, Acta Mech. \textbf{175} (2005) 77.

\bibitem{Puntigam:1996vy}
R. A. Puntigam and H. H. Soleng, Class. Quant. Grav. \textbf{14} (1997), 1129.

\bibitem{daSilva:2019lzh}
W. C. F. da Silva and K. Bakke, Eur. Phys. J. C \textbf{79} (2019) 559.

\bibitem{Vachaspati0}
T. Vachaspati, Phys. Rev. D \textbf{44} (1991) 3723.

\bibitem{Ramberg}
N. Ramberg, W. Ratzinger and P. Schwaller, JCAP \textbf{02} (2023) 039.

\bibitem{Yamada}
M. Yamada and K. Yonekura, Phys. Rev. D \textbf{106} (2022) 123515.

\bibitem{WAhmed}
W. Ahmed, M. Junaid, S. Nasri and U. Zubair, Phys. Rev. D \textbf{105} (2022) 115008.

\bibitem{HChen}
H. Chen, Z. W. Long, C. Y. Long, S. Zare and H. Hassanabadi, Int. J. Geom. Meth. Mod. Phys. \textbf{19} (2022) 2250133.

\bibitem{KBakkeH}
K. Bakke and H. Mota, Gen. Rel. Grav. \textbf{52} (2020) 97.

\bibitem{BoumaliNM}
A. Boumali and N. Messai, Can. J. Phys. \textbf{95} (2017) 999.

\bibitem{HassanabadiHM}
H. Hassanabadi, M. Hosseinpour and M. de Montigny, Eur. Phys. J. Plus \textbf{132} (2017) 541.

\bibitem{ZWang}
Z. Wang, Z. W. Long, C. Y. Long and B. Q. Wang, Can. J. Phys. \textbf{95} (2017) 331.

\bibitem{MulerK}
H. J. W. M\"{u}ler-Kirsten, Introduction to quantum mechanics:
Schrödinger equation and path integral (Word Scientific, Singapore,
2006).

\bibitem{Abramowitz}
M. Abramowitz, I.A. Stegum, Handbook of mathematical functions
(Dover Publications Inc., New York, 1965).

\bibitem{Butkov}
E. Butkov, \textit{Mathematical Physics}, (1st ed., Addison-Wesley, USA, 1973).

\bibitem{SNCF}
A. L. Silva Netto, C. Chesman and C. Furtado, Phys. Lett. A {\bf 372} (2008) 3894.

\bibitem{BBM}
I. Bialynicki-Birula and J. Mycielski, Comm. Math. Phys. {\bf 44}, (1975) 129.

\bibitem{CI}
C. O Edet and A. N. Ikot, Eur. Phys. J. C \textbf{136} (2021) 432.

\bibitem{Bezerra:1997mn}
V. B. Bezerra, J. Math. Phys. \textbf{38} (1997), 2553.

\bibitem{Aptekarevh}
A. I. Aptekarev, J. S. Dehesa, R. J. Yáñez, J. Math. Phys. {\bf 35} (1994) 4423.

\bibitem{Assch}
W. Van Assche, R. J. Yáñez, J. S. Dehesa, J. Math. Phys. {\bf 36} (1995) 4106.

\bibitem{Hirschmann}
I. I. Hirschmann Jr., Amer. J. Math. {\bf 79} (1957) 152.
\end{thebibliography}
\end{document}